# Low-Voltage Electron Emission by Graphene-hBN-graphene Heterostructure

*Zhexuan Wang, Fang Liu\*, Kaiyu Cui, Xue Feng, Wei Zhang, and Yidong Huang\**

Department of Electronic Engineering, Tsinghua University, Beijing, China; Beijing National Research Center for Information Science and Technology, Tsinghua University, Beijing, China.

ABSTRACT

Scanning Electron Microscopes (SEM) with low energy electron sources (accelerating voltage of less than 1000V) have important application requirements in many application scenarios. Tunneling junction can potentially achieve low-voltage and planar-type electron sources with good emission current density. However, further lower the extracting voltage while ensure the emission current density remains challenging. In this paper, we report a low-voltage planar-type electron source based on graphene-hBN-graphene heterostructures (GBGH) under a really low out-plane extracting voltage. The external electric field strength applied to the electron sources is only $4\times10^4$V/m and the accelerating voltage as low as 20V is realized. Steady electron emission of over 1nA and operating duration of several hours is observed from the GBGH with size of 59.29$\mu m^2$ in our experiments, and thus the maximum emission current density reaches 7mA/$cm^2$. Great electrical contacts, extremely low thickness, and excellent layer properties of two-dimensional (2D) materials lead to easy-fabrication and miniature on-chip electron sources, which would significantly contribute to the development of next-generation free electron devices.

# I. INTRODUCTION

Electron source is an important component of electron microscopy and free electron devices. Compared to the high energy free electrons generating from thermionic emission[1], cold field emission[2], and Schottky emission process[3], the low energy electron sources with accelerating voltage of less than 1000V shows desirable properties[3]. Take electron microscopy, for example, the short penetration depth into materials leads to the reduced interaction volume between electron and specimen, resulting in high spatial resolution and high image contrast with much fine surface detail[3]. Semiconductor and insulator materials can also be detected directly by low energy electrons due to the reduction of radiation damage and charging effect on specimens[4-6].

Tunneling junction structures could potentially achieve low-voltage and planar-type electron sources with good emission current density. The traditional tunneling sources are composed of bulk materials such as bottom cathodes of silicon or metal, insulators of silicon dioxide, and top conduct layers of metal[7-22]. However, severe inelastic scattering of electrons during the emission process causes bad performance of current density and efficiency[23-26]. Others reported the partial replacement of bulk materials by 2D materials such as graphene and hexagonal boron nitride (hBN) to suppress the electron-electron elastic scattering and electron-phonon inelastic scattering processes[27-34]. Yet the external electric field strength used to extract and collect emitting electrons is still relatively high, and also the emission current density remains to be improved.

In this paper, we achieve the large emission current density based on low-voltage planar-type electron sources with 2D materials of graphene-hBN-graphene heterostructure (GBGH) under a really low out-plane voltage. The electric properties of GBGH are studied numerically by first-principal calculation, where we verify the static band structure of tunneling junction and band shifts behaviors under certain electric field. The acceleration voltage applied to the electronic source is only 20V, which means that the external electric field strength is only $4\times10^4$V/m. Steady electron emission of over 1nA with operating duration of several hours is observed from 59.29μm$^2$ stacking GBGH area in our experiments, and thus the maximum detected current density is 7mA/cm$^2$. The *I-V* curve of GBGH is well explained according to the Fowler-Nordheim tunneling equation, which indicates that the emitted electrons are results from heterostructure.

## II. STRUCTURES AND NUMERICAL ANALYSIS

### A. The structure of GBGH electron source

Figure 1(a) presents the schematic description of GBGH electron sources, where 2D materials are stacked vertically together on a quartz substrate. From bottom to top, they are bottom-layer graphene, hBN, and top-layer graphene. Two layers of graphene are electrically insulated from each other by being separated with hBN.

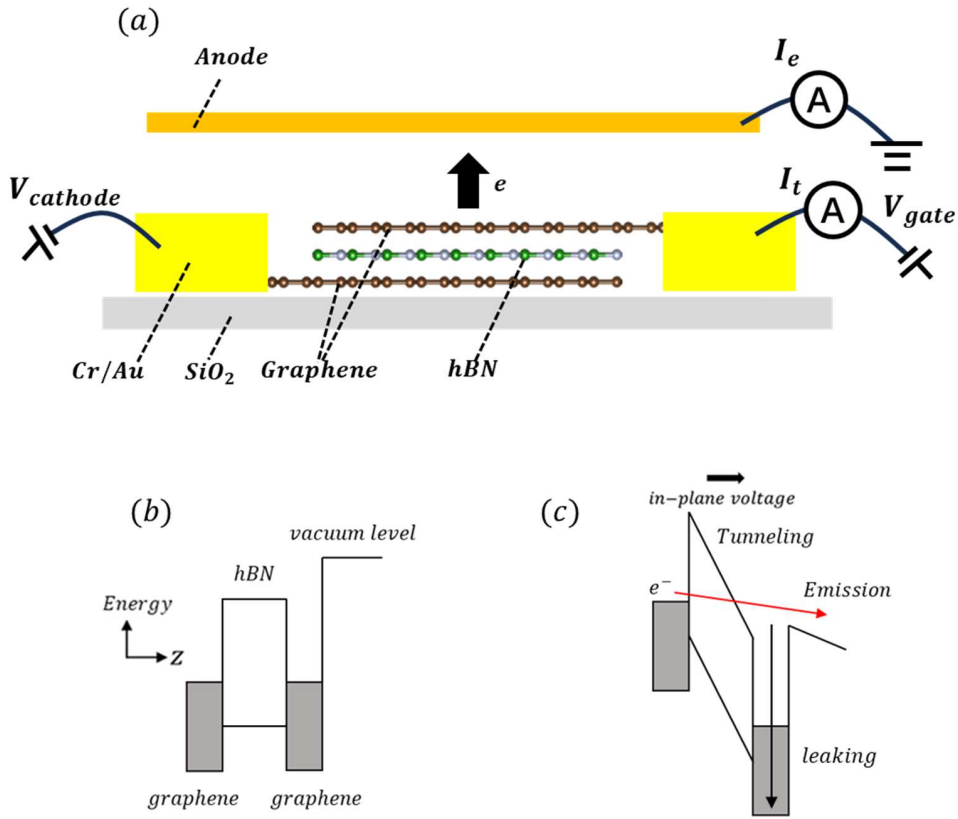

**Figure 1.** (a) The schematic of electron emitter consisting of graphene-hBN-graphene heterostructure (GBGH). Cr/Au electrodes connected to two layers of graphene respectively, which are electrically insulated from each other by being separated with hBN. (b) Without and (c) With in-plane driving voltage, the simplified schematic diagram of the energy band of graphene-hBN-graphene heterostructure. The horizontal axis represents different material levels, while the vertical axis represents electron energy.

The top and bottom graphene of GBGH is connected to 50nm Cr/Au electrodes, respectively. A fixed gate voltage $V_{gate}$ of -20V is maintained, while the cathode voltage $V_{cathode}$ is varied. Therefore, in-plane driving voltage ($V_{DR}$) is defined as the difference between $V_{gate}$ and $V_{cathode}$. By applying $V_{DR}$, the electrons emit from the stacking (heterostructure) area as planar electron source. A copper-made anode collector connected to a pico-ammeter is positioned 500μm above the substrate plane to collect emitted electrons $I_e$. As a result, the out-plane collecting electric field is only $4\times10^4$V/m. To accurately measure the leaking gate currents $I_l$, another pico-ammeter is utilized.

The schematic diagram of simplified band structure of GBGH is shown in Fig. 1(b)-(c), where the vertical axis represents energy and the *z*-axis represents the position of electrons. The white area represents hBN and exhibits insulator properties. The positions of the bottom of the conduction band and the top of the valence band for hBN are indicated by two solid black lines in the white area. Two graphene layers are exhibited by the gray area with the top black line representing the position of the Fermi level. From the relative relations of the energy bands of three layers, it can be seen that, when no $V_{DR}$ is applied (namely $V_{DR}= 0$), the Fermi level position of both graphene layers is the same. If $V_{DR}$ is applied on the electrodes, the width of the potential barrier created by hBN is significantly modified and narrowed, which causes a major shift in the Fermi levels of graphene layers. This shift leads to a considerable increase in tunneling rates for carriers from bottom-layer graphene. As carriers gain energy in the tunneling process, the required out-plane acceleration voltage for electron emission can be greatly reduced. Electrons that achieve enough energy to exceed the surface vacuum level are emitted from the top-layer graphene, as shown by the red arrow in Fig 1(c). In cases where this energy threshold is not surpassed, the carriers leak as in-plane gate current instead, as shown by the black arrow in Fig 1(c).

## B. Density functional theory (DFT) calculations

We calculate the electrical properties of GBGH under different electric field strength by Vienna Ab-initio Simulation Package (VASP) code[35]. See details in *Methods*. The heterostructure is modeled by three layers of C, B and N atoms as shown in Fig. 2(a). The supercell shape of unit cell is $\sqrt{7}\times\sqrt{7}$ R19°, where the lattice constant is optimized to 2.479Å to minimize the lattice mismatch. Vacuum layer is 30Å in the direction normal to the interface to simulate the isolated boundary.

In Fig. 2(b), the projected band structure of GBGH without interlayer electric field is presented. The corresponding energy bands from each atomic orbital of graphene and hBN is demonstrated and represented by the green and red line. The projected band of GBGH clearly shows the Dirac point of graphene (green circle) and energy band gap of hBN (red line), indicating that each layers maintain their electrical properties after forming heterostructures. The energy difference between conduct band bottom of hBN and Dirac point is calculated as 3.317eV, which is considered as the potential barrier height in tunneling junction.

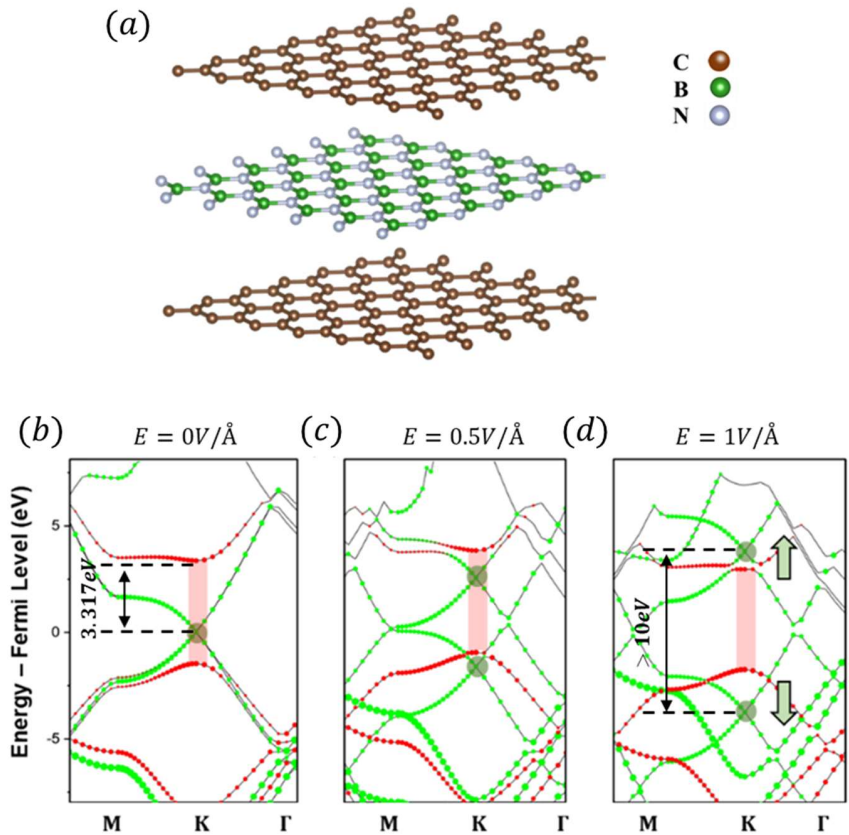

**Figure 2.** (a) The graphene-hBN-graphene heterostructure (GBGH) model in density functional theory (DFT) calculations. (b) Projected band structure of GBGH without electric field. (c) Projected band structure of GBGH with electric field of E=0.5V/Å. (d) Projected band structure of GBGH with electric field of E=1V/Å. The Dirac point of graphene (green circle) shift in opposite direction while electrical properties of each layers remain. In (b)-(d), The corresponding energy bands from each atomic orbital of graphene and hBN is represented by the green and red line.

By applying interlayer static electric fields, the projected band structure clearly changes as in Fig. 2(c)-(d). The direction of the electric field is from the top-layer to the bottom-layer graphene.

As the direction of applied fields normal to the interfaces, both Dirac points of graphene (green circle in Fig. 2(c)-(d)) clearly shift in opposite directions while electrical properties of each layers remain. With the increased electric field, the energy difference between the two Dirac points of graphene significantly increases to over 10eV under a 1V/Å of electric field, and the Dirac points of the bottom layer graphene even exceed the conduction band edge of hBN, as shown in Fig 2(d). Meanwhile, the large band gap of hBN stay unchanged. The band structure of GBGH under electric fields indicates that carriers in bottom-layer graphene have sufficient kinetic energy to tunnel or surpass the barrier of hBN. By applying a relatively low out-plane voltage, electrons are quite possible to escape from the materials. Also, GBGH is theoretically capable to sustain the electric properties under interlayer static electric fields.

## III. EXPERIMENT RESULTS
### A. Fabrications

The GBGH electron sources are prepared by mechanical exfoliation and stacking methods. See details in *Methods*. Few-layer flakes of hBN and graphene is processed. Figure 3(a)-(b) show the transmission and reflection optical lens images of the heterostructure, where the stacking area is measured as 59.29μm$^2$. Atomic Force Microscopy (AFM) is used to characterize the thickness and surface feature of heterostructure, as shown in Fig. 3(c). The thickness of few-layer hBN is 19.923nm while the top and bottom layers of graphene is 12.14nm and 26.51nm, respectively.

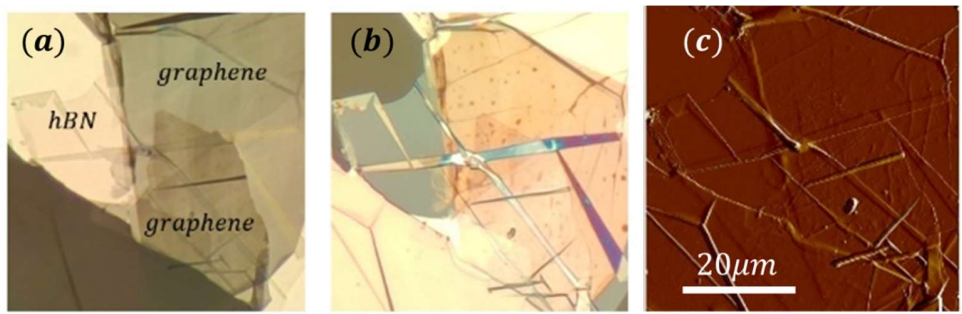

**Figure 3.** Transmission (a) and reflection (b) optical lens images of fabricated graphene-hBN-graphene heterostructure (GBGH). (c) Atomic Force Microscopy (AFM) measurements of GBGH, where the thickness of few-layer hBN is 19.923nm while the top and bottom layers of graphene is 12.14nm and 26.51nm, respectively.

## B. Measurement results

Setting $V_{DR}$ as 20V and 24V, the emission current from GBGH is well measured. It can be found from Fig. 4(a) that the stability of the emission current from GBGH is well performing at level of 0.1nA and 1nA. Steady electron emission of over 1nA and operating duration of 100s is observed from area of 59.29μm² stacking part. Therefore, the maximum detected current density is over 7mA/cm². The higher $V_{DR}$ results in larger emission current because it provides more kinetic energy. The maximum lifetime of devices is over 1 hour during the tests, which remains to be improved.

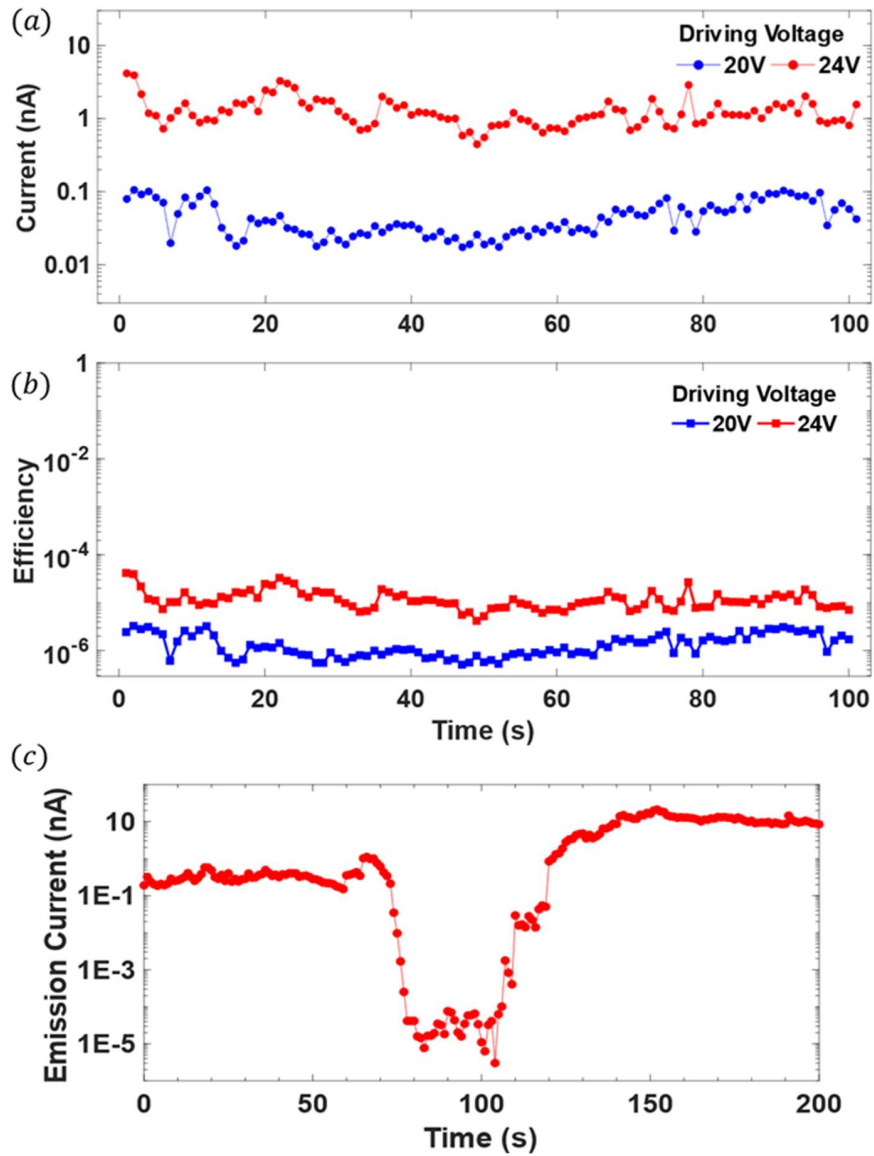

**Figure 4.** (a) Graphene-hBN-graphene heterostructure (GBGH) electron source emission current. Stable electron emission about 0.1nA and 1nA for a duration of 100 seconds. (b) The emission efficiency test results of GBGH. At 20V and 24V driving voltages, the average electron emission efficiency is less than 0.01%. (c) Switch test. During emission, when the driving voltage is turned off and restarted, the emission current will be turned off and generated simultaneously.

The emission efficiency is defined here as the ratio of the emission current to the total tunneling current. In Fig. 4(b), the emission efficiency under $V_{DR}$ of 20V and 24V are shown. An extremely low efficiency of lower than 0.01% is found in our experiments, which mainly due to the large leaking tunneling current because of the structural damage in GBGH. Specifically, the hBN layers have an electric breakdown in part of the stacking area during the test. Nonuniform layer thickness results in different breakdown electric field thresholds. Nevertheless, a higher $V_{DR}$ results in a better emission efficiency. It is due to the higher interlayer electric field, which allows more tunneling carriers to overcome the vacuum energy level of the top-layer graphene and be emitted. Therefore, more electrons emit under higher $V_{DR}$.

When $V_{DR}$ is turned off and restarted, the emission follows by the shutdown and regeneration at the same time, demonstrating a strong correlation. $V_{DR}$ shifts the barrier heights of tunneling heterostructure, allowing the electron sources to switch on and off at any time. In Fig. 4(c), the GBGH electron source shows a good switching behavior. Steady signals of 0.1nA lasts for about 50s at first. Then the signal suddenly vanishes when $V_{DR}$ shuts down. After then the emission stays at the noise level, which is around 0.1pA. We turn on the $V_{DR}$ again gradually and the emission current regenerates as expected to the previous amplitudes. The restarted emission stably lasts for at least 100s. Such behavior verifies the tunneling emission process and shows huge potentials of more flexible applications.

Further, the emission current can be adjusted by finely changing the $V_{DR}$ as well. We obtain in Fig. 5(a) the relation between average emission current to $V_{DR}$ and GBGH demonstrates well tunable emission behaviors. The emission current rises from 0.01nA to 0.1nA as the increase of $V_{DR}$ from 23V to 24.5V. Replotting Fig. 5(a) by changing the axis as $\ln(I/V^2)$ and $1/V$, we further obtain the relation between average emission current to $V_{DR}$, as shown in red circle in Fig. 5(b). For ensuring the electrons are emitted from heterostructure, we consider the Fowler-Nordheim (FN) equation, which describes the tunneling rates of electrons when penetrating through a potential barrier. The FN equation is as follows[36]

$$J = \frac{AV^2}{\phi d^2} e^{-\frac{B\phi^{1.5}V}{d}} \quad \cdots\cdots\cdots\cdots (1)$$

where $A, B$ are constants, $\phi$ is the barrier height and $d$ is the total thickness of hBN layers. It can be seen in Fig. 5(b) that $\ln(I/V^2)$ and $1/V$ satisfies a linear correlation, with a slope of $k = B\phi^{1.5}d$, from which the theoretical thickness of hBN can be calculated. Fitting the test points by linear regression analysis, as shown by the black dashed line in Fig. 5(b), we derive the linear line expression as

$$y = -935.06x + 8.8341, r^2 = 0.9416 \quad \cdots\cdots\cdots\cdots (2)$$

where the excellent fitting residual statistical properties proves the effectiveness of linear regression analysis.

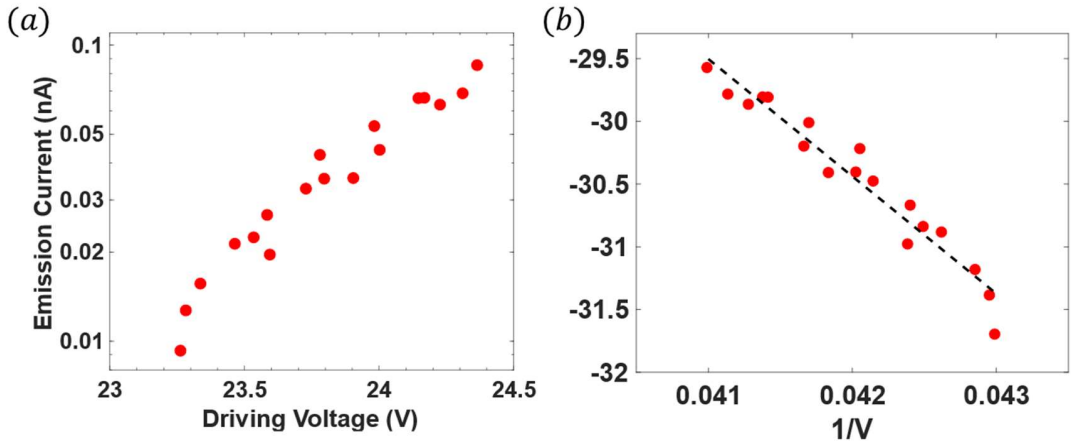

**Figure 5.** (a) Graphene-hBN-graphene voltage regulated current test. By changing the driving voltage, the emission current of the electronic source can be finely adjusted. As the driving voltage increases from 23V to 24.5V, the emission current increases from 0.01nA to 0.1nA. (b) The relation between average emission current to the driving voltage, as shown in red circle. The results fitted using the Fowler-Nordheim equation are represented by the black dashed line.

As mentioned above, the calculated barrier height of tunneling structure is 3.317eV. Based on the FN equation, the thickness of hBN extracted from the experiment curve slope in Fig. 5(b) is 22.65nm. Then we use AFM to measure the actual hBN thickness of 19.923nm. Compared the extracted thickness with the actual thickness of hBN, the emission current process is well verified by FN tunneling equation. This consistent with the extracted thickness from FN equation and well proves that the emission law conforms to the FN tunneling theory.

## IV. DISCUSSIONS

In our experiments, we first observed a substantial emission current from GBGH electron sources, at the nearly same level to the results of recently reported tunneling electron[29, 34, 37] sources. Nevertheless, the emission efficiency is not satisfactory. We consider the following factors have a decisive impact on the emission current of electronic sources.

### A. The Thinness of 2D Materials

The process of electron escape from the material involves inelastic scattering between electrons and the lattice, leading to significant energy loss[33, 38-40]. The direction of emitted carrier is perpendicular to the interfaces, which means the thinness of GBGH results in reduced inelastic scattering. Consequently, energy loss in the tunneling path is extensively suppressed, allowing more electrons to emit from the device. Reducing the thickness of the cathode material significantly lowers the energy loss of electrons within the material, thereby ensuring a higher electron emission density.

### B. Excellent Electrical Contact Between Layers

The thin-layer structure of 2D materials, with strong covalent bonds within the plane, provides material stability under low thickness. Also, fewer dangling bonds on the material's surface effectively improve electrical contact with other layers, avoiding Fermi level pinning effect or Schottky barriers. Clear interfaces with minimal contamination ensure a smooth tunneling and emission path. Moreover, materials like graphene and boron nitride, all with hexagonal lattice structures[41-43], offer better interlayer lattice matching compared to bulk materials.

### C. Large leaking tunneling current

Graphene is a zero-bandgap conductor with extremely high conductivity due to the Dirac cone in its band. In fact, tunneling junctions based on 2D material heterostructures are difficult to operate with high voltage differences or voltage fluctuations[44, 45]. It is necessary to minimize interlayer leakage current to avoid problems such as device heating caused by large tunneling currents[46]. However, compared to bulk materials, we found that the graphene layer is voltage-fluctuation intolerant and also not very robust. Therefore, the lifespan of the device is greatly limited, and the structure is easily affected by voltage fluctuations such as static electricity or power switching. The large leakage or short circuit in part of the tunneling structure greatly aggravate the tunneling current.

## V. CONCLUSIONS

In this paper, we realize a planar-type electron sources based on fully 2D materials of GBGH operated under a really low out-plane voltage. The electric properties of GBGH are studied by first-principal calculation and the existence of tunneling junction in the static band structure of GBGH is verified. By applying electric fields, energy bands shift and tunneling path appears as expected. The GBGH electron sources are prepared by mechanical exfoliation and stacking methods and driven by accelerating voltage of as low as 20V (the external electric field of only $4\times10^4$V/m). Steady electron emission of over 1nA is observed from 59.29μm$^2$ stacking area with operating duration of several hours, and the maximum current density of about 7mA/cm$^2$. The *I-V* curve of GBGH is well explained according to the FN tunneling equation indicating the electrons indeed emit from the tunneling structure. The easy-fabrication and miniature on-chip electron sources with extremely low driving voltage will establish the promising prospect of the development of next-generation free electron devices.

## VI. Methods

### A. Computational Methods

We perform all our calculations with Vienna Ab-Initio Simulation Package (VASP) code. The general gradient approximation (GGA) with the Perdew–Burke–Ernzerhof (PBE) function is utilized for the exchange-correlation energy of the interacting electrons with the projector augmented wave (PAW) potentials. We use Monkhorst-Pack 24×24×1 *k*-points grids for atomic relaxations. An energy cutoff 500eV and the electronic convergence criterion $10^{-5}$eV is set as the optimization. Atoms are allowed to be fully relaxed till the atomic forces is less than 0.01eV/Å.

### B. Device Fabrication

We choose polished quartz plate as the substrates. Bulk graphite and Boron nitride crystal is prepared. We use traditional exfoliate methods by blue tape to fabricate few-layer flakes of graphene and hBN. Polydimethylsiloxane (PDMS) of 4mm thickness is utilized to transfer the chosen flakes to the substrates. PDMS become non-stick and easy to be removed after baking at 80°$C$ for at most 20 min. The stacking flakes on substrates are annealed at 300°$C$ for 1h. Then, gold electrodes are prepared by ultraviolet lithography with NR9-3000PY resist and deposition of Cr/Au (5nm/20nm) by magnetic sputter method.

### C. Experiment Characterization

We build a homemade vacuum chamber to create the high vacuum test environment. All the devices are measured at a pressure of $10^{-8}$torr. Emission current and in-plane gate current are measured by pico-ammeters of Keithley 6485 and 6487, respectively. Atomic forces microscope of BRUKER Dimension Fast Scan system is used to characterize the thickness and morphological image of devices. Optical microscope of Nikon Eclipse LV100D system is also performed.

AUTHOR INFORMATION

**Corresponding Authors**

* liu_fang@tsinghua.edu.cn; yidonghuang@tsinghua.edu.cn

**Author Contributions**

All authors have given approval to the final version of the manuscript.

ACKNOWLEDGMENT

This work was supported by the National Key Research and Development Program of China (2023YFB2806703) and the National Natural Science Foundation of China (Grant No. U22A6004).